# EFFECTIVE ELASTIC PROPERTIES OF PLANAR SOFCs: A NON-LOCAL DYNAMIC HOMOGENIZATION APPROACH

A. Bacigalupo[1], L. Morini and A. Piccolroaz

Department of Civil, Environmental and Mechanical Engineering,

University of Trento, via Mesiano, 77, I-38123 Trento, Italy

**Abstract**

The focus of the article is on the analysis of effective elastic properties of planar Solid Oxide Fuell Cell (SOFC) devices. An ideal periodic multi-layered composite (SOFC-like) reproducing the overall properties of multi-layer SOFC devices is defined. Adopting a non-local dynamic homogenization method, explicit expressions for overall elastic moduli and inertial terms of this material are derived in terms of micro-fluctuation functions. These micro-fluctuation function are then obtained solving the *cell problems* by means of finite element techniques. The effects of the temperature variation on overall elastic and inertial properties of the fuel cells are studied. Dispersion relations for acoustic waves in SOFC-like multilayered materials are derived as functions of the overall constants, and the results obtained by the proposed computational homogenization approach are compared with those provided by rigorous Floquet-Boch theory. Finally, the influence of the temperature and of the elastic properties variation on the Bloch spectrum is investigated.

---

[1] Corresponding author: andrea.bacigalupo@unitn.it

# 1. Introduction

Fuell cells are one of most promising small-scale power generation systems which can play an important role in meeting the increasing global demand of renewable energy. These devices convert the chemical energy of fuels directly into electrical energy achieving an high energy conversion efficiency (50-60%).

In solid oxide fuel cells (SOFC), a layer of doped ceramic materials is used as electrolyte. The solid-state electrolyte is sandwiched between two high-temperature endurable porous-media electrodes (Bove and Ubertini, 2008). The choice of the materials for these components depends on the operating temperature, typically included in the range 600–1000°C (Pitakthapanaphong and Busso, 2005), and a detailed review of the most commonly used options in SOFC fabrication and their operative implications is reported in Zhu and Deevi, 2003, Brandon and Brett, 2006.

The operating principle of SOFC devices is analogous to that of lithium ions battery (Hajimolana *et al.* 2011): the oxidant (oxygen) is supplied to the cathode, where it receives electrons from an extern electrical circuit and then reduces to oxygen ions. The oxygen ions pass through the electrolyte and reach the anode surface where they combine with the $H_2$ fuel and form water. The recombination reactions generate electrons which travel through the circuit providing electrical current. The single cells composed by anode, electrolyte and cathode are provided of conducting interconnections on each side (see Figure 1.a) and stacked together to produce enough voltage for practical use. A scheme of multi-layer SOFC device is reported in Figure 1.b.

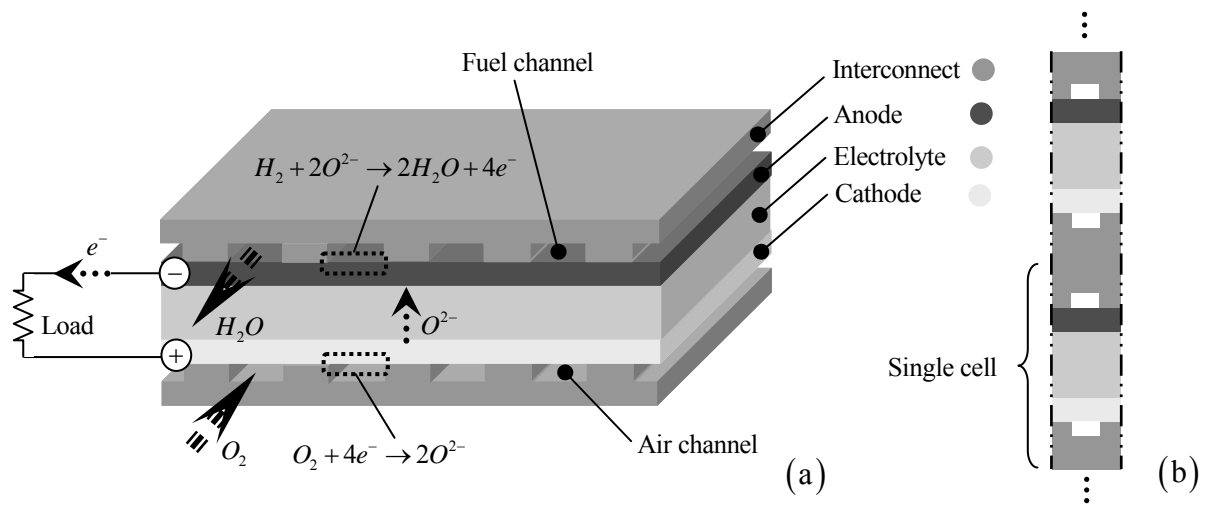

Figure 1. (a) planar solid oxide fuel cell (SOFC); (b) Schematic multi-layer SOFC device.

Due to the high operating temperature, the SOFC components are subject to severe thermomechanical stresses which can cause damage and crack formation compromising the performance of the devices in terms of power generation and energy conversion efficiency (Atkinson and Sun, 2007, Kuebler *et al.*, 2010). Modelling the mechanical properties of solid oxide fuell cells represent a crucial issue in order to predict these phenomena and then to ensure the successful manufacture and reliability of the system. Despite this, most of the work in this field has been concentrated on the electrochemical rather than the mechanical aspects of the SOFC materials (Kakac *et al.*, 2007, Colpan *et al.*, 2008). Relatively few studies on the mechanical behaviour and failures of the SOFC components have been performed. Most of them regard the mechanical properties of individual materials that constitute the SOFC rather than the overall behaviour of the multi-layered structure. To the author's knowledge, a rigorous quantitative study regarding the overall elastic and inertial properties of the multi-layered SOFC is still unknown in literature.

The principal aim of this paper is to provide exact analytical expressions to estimate the overall elastic moduli and inertial terms of multi-layered SOFC devices. With this purpose, an ideal periodic multi-layered composite (SOFC-like) reproducing the overall properties of multi-layer SOFC reported in Figure1 is introduced in the paper.

The static and dynamic properties of composite materials having periodic microstructure, such as SOFC-like media here introduced, have been intensively studied by means of multi-scale approaches (Kim et al., 2009, Salvadori et al., 2014). Indeed, in order to obtain a synthetic mechanical description of periodic materials it is necessary to consider different scales of observation introducing homogenization techniques. These approaches associate to the considered heterogeneous material at the micro-scale, described by a standard Cauchy continuum, an equivalent homogenous medium at the macro-scale. The behaviour of the equivalent macroscopic material can be described by means of a first order continuum or alternatively a non-local medium (see Bakhvalov and Panasenko, 1984).

In the cases where the size of the microstructural components is not negligible with respect to the structural dimension, as for the considered SOFC-like periodic medium, the standard approach based on first order homogenization may present some disadvantages. Generalized continua such as multipolar (second gradient, e.g. Smyshlyaev and Cherednichenko, 2000, Bacigalupo and Gambarotta, 2013a, Bacca *et al.* 2013a,b,c, among the others), micromorphic (Cosserat, e.g. Forest and Sab 1998, Forest and Trinh, 2011, De Bellis and Adessi, 2011, Adessi *et al.* 2013, among the others) and structured deformations (Deseri and Owen, 2003, 2010) continua may provide better results by introducing characteristic lengths in the constitutive model. Moreover, spatial and temporal multiscale asymptotic homogenization techniques can be successfully used to perform both static and dynamic thermo-mechanical analysis in heterogeneous multiphase materials (Zhang *et al.*, 2007, Kanoute et al., 2009, Schrefler *et al.*, 2011).

In order to estimate overall elastic moduli together with global inertial constants, in this work the introduced periodic SOFC-like material is treated by means of the non-local dynamic homogenization approach developed in Bacigalupo and Gambarotta, 2013b and Bacigalupo, 2014. This method allows us to derive the exact expressions for overall elastic and inertial constants taking into account characteristic size effects and dispersion induced by microstructure.

The homogenization results are used to study the effects of the temperature variation on overall elastic and inertial properties of the SOFC. These effects are of particular relevance in order to predict the behaviour of the devices subject to both room temperature and high operative temperature. Moreover, analytic dispersion functions for the acoustic waves are determined using the overall constants derived by homogenization procedure, and the effect of the temperature on waves propagation is also analyzed. The obtained results have been compared with those provided by rigorous Floquet-Bloch theory; a few notes are reported in Appendix B of the paper. Floquet-Bloch approach is also adopted to estimate the influence of the temperature on the high frequency propagation branches which cannot be determined by means of standard homogenization procedures. The variation of the Bloch spectrum caused by lacking of centro-symmetry in the periodic cell due to different elastic properties of the anode and the cathode is analyzed and its physical implications are discussed in details.

The article is organized as follows: in Section 2, the geometry of the idealized periodic SOFC-like material is illustrated, the local equation of motion describing the micro-displacement, micro-strain and micro-stress fields is introduced., and the macroscopic quantities characterizing the second gradient equivalent continuum are defined.

Exact expressions for the overall elastic moduli and inertial terms are obtaind by means of dynamic homogenization procedure in Section 3. In the same Section, analytical dispersion relations for acoustic waves in the equivalent second gradient material are derived. The acoustic wave propagation in SOFC-like materials is studied in Section 4. As just anticipated, the results obtained using the homogenization approach is tested by means of a comparison with Floquet-Bloch theory. Finally, a critical discussion about the obtained results is reported together with conclusions in Section 5.

## 2. Multi-scale modelling of periodic materials

Let us consider a multi-layered composite (planar SOFC-like) having periodic micro-structure as shown in Figure 2.a and subject to small strains; each of its constituent elements is modelled as an elastic Cauchy continuum (non-linear constitutive equations may be considered if incremental approaches are adopted). The material point is identified by vector position $\mathbf{x} = x_1\mathbf{e}_1 + x_2\mathbf{e}_2$ referred to a system of coordinates with origin at point O and orthogonal base $(\mathbf{e}_1, \mathbf{e}_2)$. The periodic material is fully characterized by the periodic cell $\mathcal{A} = [0,\varepsilon] \times [0,\delta\varepsilon]$ with characteristic size $\varepsilon$ shown in Figure 2.b, which is spanned by the two independent orthogonal vectors $\mathbf{v}_1 = d_1\mathbf{e}_1 = \varepsilon\mathbf{e}_1$, $\mathbf{v}_2 = d_2\mathbf{e}_2 = \delta\varepsilon\mathbf{e}_2$. Accordingly, $\mathcal{A}$ is the elementary cell period of both the micro-elasticity tensor $\mathbb{C}^{m,\varepsilon}(\mathbf{x})$ and the mass density $\rho^\varepsilon(\mathbf{x})$ with defined as: i.e. $\mathbb{C}^{m,\varepsilon}(\mathbf{x}+\mathbf{v}_i) = \mathbb{C}^{m,\varepsilon}(\mathbf{x})$, $\rho^\varepsilon(\mathbf{x}+\mathbf{v}_i) = \rho^\varepsilon(\mathbf{x})$, $i$=1,2, $\forall \mathbf{x} \in \mathcal{A}$ (they are commonly denoted as $\mathcal{A}$-periodic function). This suggests to consider a unit cell $Q = [0,1] \times [0,\delta]$ that reproduces the periodic microstructure by rescaling with the small parameter $\varepsilon$ such that the two distinct scales are represented by the macroscopic (slow) variables $\mathbf{x} \in \mathcal{A}$ and the microscopic (fast) variable $\xi = \mathbf{x}/\varepsilon \in Q$. The mapping of both the elasticity tensor and of the mass density may be defined on $Q$ as follows: $\mathbb{C}^{m,\varepsilon}(\mathbf{x}) = \mathbb{C}^m(\xi = \mathbf{x}/\varepsilon)$, $\rho^\varepsilon(\mathbf{x}) = \rho(\xi = \mathbf{x}/\varepsilon)$, respectively.

The micro-displacement $\mathbf{u}(\mathbf{x},t)$ is considered together with the corresponding micro-strain tensor $\boldsymbol{\varepsilon}(\mathbf{x},t) = sym\nabla \mathbf{u}(\mathbf{x},t)$ and the micro-stress tensor $\boldsymbol{\sigma}(\mathbf{x},t) = \mathbb{C}^m\left(\dfrac{\mathbf{x}}{\varepsilon}\right)\boldsymbol{\varepsilon}(\mathbf{x},t)$ which has to satisfy the local equation of motion $\nabla \bullet \boldsymbol{\sigma}(\mathbf{x},t) = \rho\left(\dfrac{\mathbf{x}}{\varepsilon}\right)\ddot{\mathbf{u}}(\mathbf{x},t) - \mathbf{f}(\mathbf{x},t)$ where $\mathbf{f}(\mathbf{x},t)$ is the body force depending on the slow variable. The resulting set of partial differential equations is written in the form

$$\nabla \cdot \left( \mathbb{C}^m \left( \frac{\mathbf{x}}{\varepsilon} \right) \nabla \mathbf{u}(\mathbf{x},t) \right) = \rho \left( \frac{\mathbf{x}}{\varepsilon} \right) \ddot{\mathbf{u}}(\mathbf{x},t) - \mathbf{f}(\mathbf{x},t) \quad , \tag{1}$$

and then the displacement may be seen in the classical form $\mathbf{u}\left(\mathbf{x}, \boldsymbol{\xi} = \frac{\mathbf{x}}{\varepsilon}, t\right)$ as a function of both the slow and the fast variable (the elasticity tensor has the property $\mathbb{C}^m \mathbf{Z} = \mathbb{C}^m sym\mathbf{Z}, \ \forall \mathbf{Z}$).

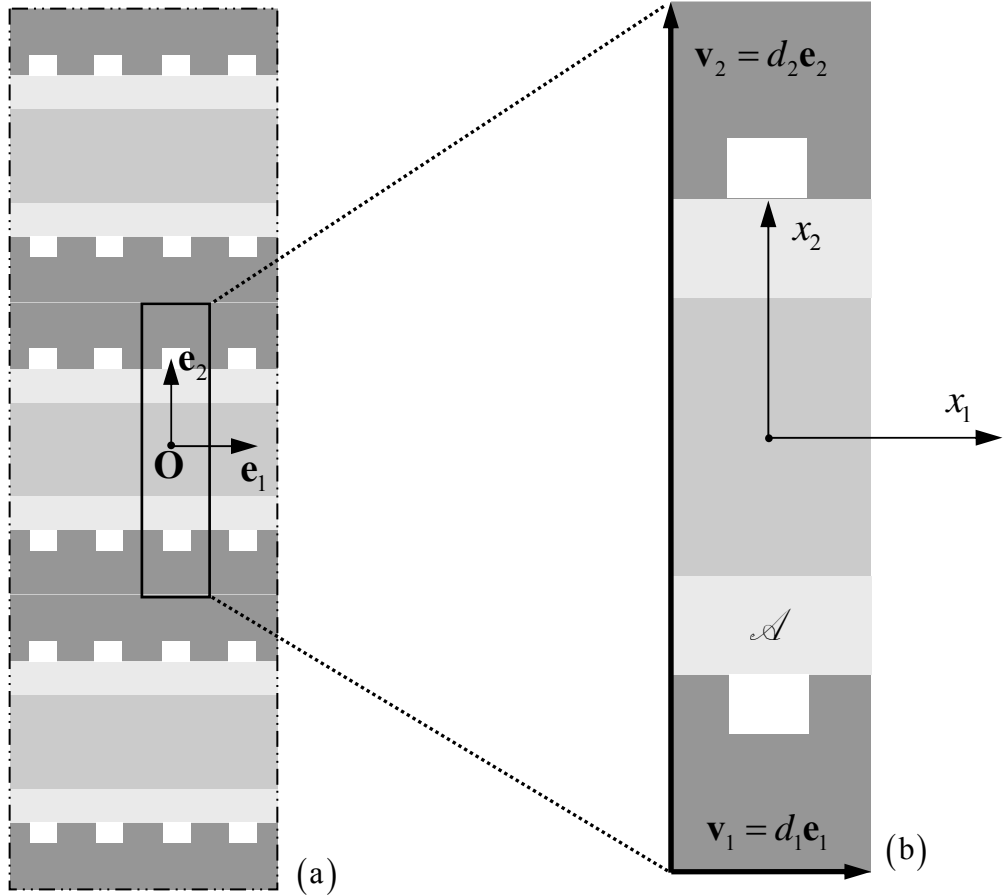

Figure 2. (a) planar SOFC-like with periodic micro-structure; (b) Periodic cell $\mathscr{A}$ and periodicity vectors.

The solution of this fine-scale problem is computationally very expensive and provides too detailed results to be of practical use, so that it is convenient to replace the heterogeneous model with an equivalent homogeneous one to obtain equations whose coefficients are not rapidly oscillating while their solutions are close to those of the original equations. In general, an equivalent classical (Cauchy) continuum is considered, that provides good results when the size of the microstructure is negligible in comparison to the structural size. However, in cases where the absolute size of the microstructure has a relevant influence on the mechanical behavior it is convenient to introduce an equivalent non-local continuum such as for example second gradient or micromorphic continua. The overall elastic moduli and inertial terms of a homogeneous second gradient continuum equivalent to periodic fuel cell reported in Figure 2 are derived following the approach recently proposed by Bacigalupo and Gambarotta, 2013b and Bacigalupo, 2014. Following this method, the overall elastic and inertial proprieties of the second gradient continuum are expressed in terms of geometrical and mechanical properties of the microstructure by means of an asymptotic expansion for the micro-displacement. The asymptotic expansion is performed in terms of the parameter $\varepsilon$, that keeps the dependence on the slow variable $\mathbf{x}$ separate from the fast one $\boldsymbol{\xi}$ such that two distinct scales are represented (Bakhvalov and Panasenko, 1984).

Considering the second gradient continuum (Mindlin, 1964) the macro-displacement $\mathbf{U}(\mathbf{x})$ of component $U_i$ is defined at point $\mathbf{x}$ in the reference $(\mathbf{e}_i, i=1,2)$ together with the displacement gradient $\mathbf{H}(\mathbf{x}) = H_{ij}\mathbf{e}_i \otimes \mathbf{e}_j = \nabla\mathbf{U}(\mathbf{x}) = \frac{\partial U_i}{\partial x_j}\mathbf{e}_i \otimes \mathbf{e}_j$ and the second-order strain $\boldsymbol{\kappa}(\mathbf{x}) = \kappa_{ijk}\mathbf{e}_i \otimes \mathbf{e}_j \otimes \mathbf{e}_k = \nabla \otimes \nabla\mathbf{U}(\mathbf{x}) = \frac{\partial^2 U_i}{\partial x_j \partial x_k}\mathbf{e}_i \otimes \mathbf{e}_j \otimes \mathbf{e}_k$. As usual, the displacement gradient can be split into the symmetric and the skew-symmetric parts $\mathbf{H}(\mathbf{x}) = \mathbf{E}(\mathbf{x}) + \boldsymbol{\Omega}(\mathbf{x})$, where $\mathbf{E}(\mathbf{x}) = sym\nabla\mathbf{U}(\mathbf{x})$ is the first-order strain, and $\boldsymbol{\Omega}(\mathbf{x}) = skw\nabla\mathbf{U}(\mathbf{x})$ is the main rotation tensor. The stress is described by the first-order stress $\boldsymbol{\Sigma}(\mathbf{x}) = \Sigma_{ij}\mathbf{e}_i \otimes \mathbf{e}_j, (\Sigma_{ij} = \Sigma_{ji})$ by the second-order stress tensor $\boldsymbol{\mu}(\mathbf{x}) = \mu_{ijk}\mathbf{e}_i \otimes \mathbf{e}_j \otimes \mathbf{e}_k, (\mu_{ijk} = \mu_{ikj})$ and by the non-symmetric real stress tensor $\mathbf{T}(\mathbf{x}) = \boldsymbol{\Sigma}(\mathbf{x}) - \nabla \bullet \boldsymbol{\mu}(\mathbf{x})$.

## 3. Second order dynamic homogenization

In this Section, elastic and inertial constants of the homogeneous second gradient continuum equivalent to a planar oxide fuel cell with periodic microstructure reported in Figure 1 are derived in explicit form by means of second order homogenization approach recently developed by Bacigalupo and Gambarotta, 2013b and Bacigalupo, 2014. The derived overall elastic moduli and inertial terms are then used in order to obtain analytic expressions for dispersion relations for acoustic waves.

### 3.1 Asymptotic analysis of micro-displacement field

The micro-displacement field is represented in the classical form $\mathbf{u}(\mathbf{x},t) = \mathbf{u}\left(\mathbf{x},\boldsymbol{\xi} = \dfrac{\mathbf{x}}{\varepsilon}, t\right)$ considered in asymptotic homogenization, where $\mathbf{x}$ and $\boldsymbol{\xi}$ play the role of slow (macro) and fast (micro) coordinates. According to Bacigalupo and Gambarotta, 2013b and Bacigalupo, 2014 the displacement may be approximated by the truncated second order asymptotic series as follows

$$\mathbf{u}\left(\mathbf{x},\boldsymbol{\xi} = \dfrac{\mathbf{x}}{\varepsilon}, t\right) \approx \mathbf{u}^{II}\left(\mathbf{x},\boldsymbol{\xi} = \dfrac{\mathbf{x}}{\varepsilon}, t\right) = \mathbf{U}(\mathbf{x},t) + \varepsilon \mathbf{N}^1(\boldsymbol{\xi}) : \mathbf{H}(\mathbf{x},t) + \varepsilon^2 \mathbf{N}^2(\boldsymbol{\xi}) \vdots \boldsymbol{\kappa}(\mathbf{x},t), \quad (2)$$

that is the superposition of the macro-displacement $\mathbf{U}(\mathbf{x},t)$, and a complementary displacement field representing the micro-fluctuation due to the material heterogeneities. The complementary field depends on the first and second gradient of the macro-displacement field $\mathbf{H}(\mathbf{x}) = \nabla \mathbf{U}(\mathbf{x})$, $\boldsymbol{\kappa}(\mathbf{x}) = \nabla \otimes \nabla \mathbf{U}(\mathbf{x})$, respectively, through the micro-fluctuation functions $\mathbf{N}^1(\boldsymbol{\xi})$ and $\mathbf{N}^2(\boldsymbol{\xi})$ (with components $N^1_{ipq_1}(\boldsymbol{\xi})$ and $N^2_{ipq_1q_2}(\boldsymbol{\xi})$).

The micro-fluctuation functions here introduced depend on the fast coordinates and are $Q$-periodic with zero mean over $Q$, namely $\langle N^1_{ipq_1}(\boldsymbol{\xi})\rangle = 0$ and $\langle N^2_{ipq_1q_2}(\boldsymbol{\xi})\rangle = 0$, where $\langle \cdot \rangle$ denotes the averaging operator over $Q$ (*normalize condition*). These unknown functions $N^1_{ipq_1}$, $N^2_{ipq_1q_2}$ are given by the solution of the following non-homogeneous equations (so-called *cell problems*, shown in Bacigalupo, 2014)

$$\begin{aligned}
\left(C^m_{tlis} N^1_{ipq_1,s}\right)_{,l} &= -C^m_{tlpq_1,l} + \tilde{h}_{tpq_1} \doteq -f^1_{tpq_1} \\
\left(C^m_{tlis} N^2_{ipq_1q_2,s}\right)_{,l} &= -\dfrac{1}{2}\Big[\left(C^m_{tliq_2} N^1_{ipq_1}\right)_{,l} + C^m_{tq_2is} N^1_{ipq_1,s} + C^m_{tq_2pq_1} + \left(C^m_{tliq_1} N^1_{ipq_2}\right)_{,l} + C^m_{tq_1is} N^1_{ipq_2,s} + \\
&\quad + C^m_{tq_1pq_2}\Big] + \tilde{h}_{tq_2pq_1} \doteq -f^2_{tq_2pq_1}
\end{aligned} \quad (3)$$

where the constants $\tilde{h}_{tpq_1}$ and $\tilde{h}_{tq_2pq_1}$ are defined as follows

$$\tilde{h}_{tpq_1} = 0, \qquad \tilde{h}_{tq_2pq_1} = \dfrac{1}{2}\left\langle C^m_{tq_2pq_1} + C^m_{tq_2is} N^1_{ipq_1,s} + C^m_{tq_1pq_2} + C^m_{tq_1is} N^1_{ipq_2,s} \right\rangle, \quad (4)$$

and the comma notation is used to indicate differentiation with respect to the *fast* variable $\xi$. According to the asymptotic approach proposed by Bakhvalov and Panasenko, 1984 the constants $\tilde{h}_{tpq_1}$ and $\tilde{h}_{tq_2pq_1}$ are determined by imposing the auxiliary body forces $f^1_{tpq}$ and $f^2_{trpq}$ (in equation (3)) possessing vanishing mean value over the unit cell $Q$, i.e. $\langle f^1_{tpq} \rangle = 0$ and $\langle f^2_{tpqr} \rangle = 0$. This assumption guarantees the $Q$-periodicity of the perturbation functions $N^j_{ipq}$ obtained as solution of problem (3) and implies the continuity of the micro-displacement fields and the anti-periodicity of the traction at the interface of adjacent cells.

The two PDEs shown in (3) may be considered as standard elasto-static problems with prescribed pseudo-body forces $f^1_{tpq}$ and $f^2_{trpq}$ that depend on the elastic moduli and on the micro-fluctuation functions. In general, the solution of both the cell problems is carried out through a FE approach with periodic boundary conditions on the micro-fluctuation functions prescribed. In the case where the mismatch in elastic moduli across different phases affects the right hand side of equation (3), this effect may be taken into account through pseudo-forces as Dirac-delta functions centrate at the interfaces.

### 3.2 Overall proprieties in homogeneous second gradient continuum

A combination of the variational and asymptotic techniques applied at the weak form of equation (1) is used to derive the second gradient homogenized equations and therefore the overall elastic moduli and inertial terms of the equivalent homogeneous material. The variational problem corresponding to equation (1) is obtained by the Hamilton's principle

$$\min_{\mathbf{u}(\mathbf{x},t)} \int_{t_0}^{t_1} \mathcal{L}(\mathbf{u}) dt = \min_{\mathbf{u}(\mathbf{x},t)} \int_{t_0}^{t_1} \left( \mathrm{T}(\mathbf{u}) - \Pi(\mathbf{u}) \right) dt = \min_{\mathbf{u}(\mathbf{x},t)} \int_{t_0}^{t_1} \int_{\mathbb{R}^2} \left( \frac{1}{2} \rho \dot{u}_i \dot{u}_i - \frac{1}{2} \frac{\partial u_i}{\partial x_j} C^m_{ijhk} \frac{\partial u_h}{\partial x_k} + f_i u_i \right) d\mathbf{x} dt \quad (5)$$

where $\mathcal{L}$ is the Lagrangian functional, T is the kinetic and $\Pi$ the potential energy functional, respectively. The averaged Lagrangian functional with respect to the parameter $\zeta \in Q$ is introduced in the form

$$\bar{\mathcal{L}}(\mathbf{u}) = \langle \mathcal{L}(\mathbf{u}) \rangle = \frac{1}{\delta} \int_Q \mathcal{L}\left( \mathbf{u}\left( \mathbf{x}, \frac{\mathbf{x}}{\varepsilon} + \zeta, t \right) \right) d\zeta = \frac{1}{\delta} \int_Q \int_{\mathbb{R}^2} \left( \frac{1}{2} \rho \dot{u}_i \dot{u}_i - \frac{1}{2} \frac{\partial u_i}{\partial x_j} C^m_{ijhk} \frac{\partial u_h}{\partial x_k} + f_i u_i \right) d\mathbf{x} d\zeta =$$
$$= \frac{1}{\delta} \int_{\mathbb{R}^2} \int_Q \left( \frac{1}{2} \rho \dot{u}_i \dot{u}_i - \frac{1}{2} \frac{\partial u_i}{\partial x_j} C^m_{ijhk} \frac{\partial u_h}{\partial x_k} + f_i u_i \right) d\zeta d\mathbf{x} \quad . \quad (6)$$

Note that expression (6) is derived assuming that the precise "phase" of the microstructure with respect to the body force is generally unknown and a family of translated microstructures should therefore be considered (see also Smyshlyaev and Cherednichenko, 2000).

If the displacement is restricted to the class of functions $\mathbf{u}^{II}$ having the form (2) and the Hamilton's principle is applied at the averaged Lagrangian functional (6), the minimization problem is written in the form

$$\min_{\mathbf{U}(\mathbf{x},t)} \int_{t_0}^{t_1} \overline{\mathcal{L}}\left(\mathbf{u}^{II}(\mathbf{U})\right) dt = \min_{\mathbf{U}(\mathbf{x},t)} \int_{t_0}^{t_1} \int_{\mathscr{L}} \left\langle \frac{1}{2}\rho \dot{u}_i^{II} \dot{u}_i^{II} - \frac{1}{2}\frac{\partial u_i^{II}}{\partial x_j} C_{ijhk}^m \frac{\partial \dot{u}_h^{II}}{\partial x_k} + f_i u_i^{II} \right\rangle d\mathbf{x}dt, \qquad (7)$$

where the averaged functional depends on the macro-displacement $\mathbf{U}(\mathbf{x},t)$.

The Euler-Lagrangian equation, associated to the variational problem (7), is truncated at the third order in the static part and to the fifth order in the inertial part (see Bacigalupo and Gambarotta 2013b). This approach is in analogy with the idea proposed in Wang and Sun, 2002 and Sun and Huang, 2007, where higher order terms were considered to obtain an accurate description of the micro-inertia effects while the constitutive parameters were assumed within the classical (first-order) description. Accordingly, in the present approach the inertial terms are defined considering the contribution of the micro-accelerations associated with the second-order strain $\boldsymbol{\kappa}(\mathbf{x},t)$. The resulting Euler-Lagrangian takes the form

$$\varepsilon^2 h_{ijkpqr} U_{i,jkqr} + \frac{\varepsilon}{2}\left(h_{ijpqr} - h_{pqijr}\right)U_{i,jqr} - h_{ijpq} U_{i,jq} = f_p(\mathbf{x},t) - \rho_M \ddot{U}_p + \varepsilon \rho_M I_{ipj} \ddot{U}_{i,j} + \\ + \varepsilon^2 \rho_M I_{iqpj} \ddot{U}_{i,qj} + \varepsilon^3 \rho_M I_{iqpjs} \ddot{U}_{i,qjs} - \varepsilon^4 \rho_M I_{iqjpsr} \ddot{U}_{i,qjsr}, \qquad (8)$$

where $h_{ijpq} = C_{ijpq}$, $h_{ijpqr} = \varepsilon^{-1} Y_{ijpqr}$, $h_{ijkpqr} = \varepsilon^{-2} S_{ijkpqr}$ and $\rho_M$, $I_{ipj}$, $I_{iqpj}$, $I_{iqpjs}$, $I_{iqjpsr}$ are the effective elastic moduli and the effective inertia parameters of the second gradient continuum, respectively and the comma derivative notation is referred to the slow variable. The overall elastic moduli defined through equation (8) have the following form

$$C_{pqrs} = \left\langle C_{ijkl}^m B_{ijpq}^H B_{klrs}^H \right\rangle, \quad \frac{Y_{pqrst}}{\varepsilon} = \left\langle C_{ijkl}^m B_{ijpq}^H B_{klrst}^\kappa \right\rangle, \quad \frac{S_{pqhrst}}{\varepsilon^2} = \left\langle C_{ijkl}^m B_{klrst}^\kappa B_{ijpqh}^\kappa \right\rangle - \frac{\left\langle A_{pqhrst}^{H-\kappa} \right\rangle}{12}, \qquad (9)$$

Moreover, the effective inertia parameters in equation (8) have the form

$$\rho_M = \langle \rho \rangle, \quad I_{ipj} = \frac{\langle \rho(N^1_{ipj} - N^1_{pij}) \rangle}{\rho_M}, \quad I_{iqpj} = \frac{\left\langle \rho\left(\frac{1}{2}(N^1_{rpj}N^1_{riq} + N^1_{rij}N^1_{rpq}) - N^2_{ipqj} - N^2_{piqj}\right)\right\rangle}{\rho_M},$$

$$I_{iqpjs} = \frac{\langle \rho(N^1_{kip}N^2_{kqjs} - N^1_{kqp}N^2_{kijs} + 2N^1_{kij}N^2_{kqps} - 2N^1_{kqj}N^2_{kips} + N^1_{kis}N^2_{kqjp} - N^1_{kqs}N^2_{kijp}) \rangle}{4\rho_M}, \quad (10)$$

$$I_{ijqpsr} = \left\langle \rho\left(N^2_{kijq}N^2_{kpsr} + N^2_{kpjq}N^2_{kisr} + N^2_{kisq}N^2_{kpjr} + N^2_{kpsq}N^2_{kijr} + N^2_{kijs}N^2_{kpqr} \right.\right.$$
$$\left.\left. + N^2_{kpjs}N^2_{kiqr} + N^2_{kijr}N^2_{kpsq} + N^2_{kirs}N^2_{kpjq} \right)\right\rangle / 8\rho_M.$$

In case of centro-symmetric periodic cell one obtains $I_{ipj} = 0$, $I_{iqpjs} = 0$. If the density is homogeneous in the unit cell, the inertial terms take the form

$$\rho_M = \rho, \quad I_{iqpj} = \frac{1}{2}\langle N^1_{rpj}N^1_{riq} + N^1_{rij}N^1_{rpq} \rangle,$$

$$I_{iqjpsr} = \frac{1}{8}\langle N^2_{kipq}N^2_{kjsr} + N^2_{kjpq}N^2_{kisr} + N^2_{kisq}N^2_{kjpr} + N^2_{kjsq}N^2_{kipr} + N^2_{kips}N^2_{kjqr} + \quad (11)$$
$$+ N^2_{kjps}N^2_{kiqr} + N^2_{kipr}N^2_{kjsq} + N^2_{kirs}N^2_{kjpq} \rangle.$$

It is important to note that if the material is homogeneous in the periodic cell, i.e. the microstructure disappears, the functions $N^1_{ikl}(\xi)$ and $N^2_{iklp}(\xi)$ are zero and both the elastic moduli and the inertial parameters defined by equations (9), (10), (11), (18) vanish and consequently the equation of motion of the classical continuum is obtained.

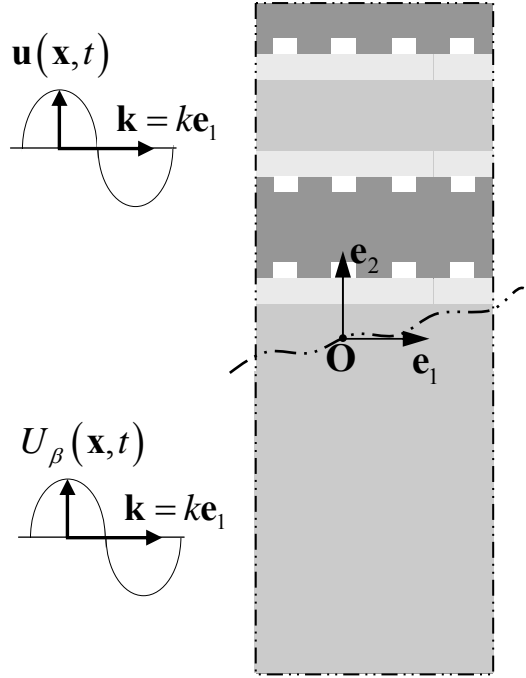

Figure 3: Wave propagation along the the orthotropy direction $\mathbf{e}_1$.

### 3.3 Dispersion relation of acoustic waves

Considering the rather common case of centro-symmetric periodic cell having orthotropic material phases (see for example Figure 3), the elastic wave propagation along the orthotropy direction $\mathbf{e}_\beta$ ($\beta = 1,2$) is described by the non zero components of the macro-displacement vector $U_\alpha(x_\beta, t)$ ($\alpha = 1,2$) (Figure 3) which are solution of the PDE

$$C_{\beta\alpha\beta\alpha} U_{\alpha,\beta\beta} - S_{\alpha\beta\beta\alpha\beta\beta} U_{\alpha,\beta\beta\beta\beta} = \rho_M \ddot{U}_\alpha - J_{\alpha\beta\alpha\beta} \ddot{U}_{\alpha,\beta\beta} + J_{\alpha\beta\beta\alpha\beta\beta} \ddot{U}_{\alpha,\beta\beta\beta\beta}, \qquad (12)$$

where indices $\alpha$, $\beta$ are not summed. Equation of motion (12) at the macro-scale can be rewritten in the analogous form

$$\left(\lambda_\beta^\alpha \hat{c}_\beta^\alpha\right)^2 U_{\alpha,\beta\beta\beta\beta} - \hat{c}_\beta^{\alpha 2} U_{\alpha,\beta\beta} = \ddot{U}_\alpha - I_{\alpha\beta\alpha\beta} \varepsilon^2 \ddot{U}_{\alpha,\beta\beta} + I_{\alpha\beta\beta\alpha\beta\beta} \varepsilon^4 \ddot{U}_{\alpha,\beta\beta\beta\beta}, \qquad (13)$$

with $\hat{c}_\beta^\alpha = \sqrt{C_{\beta\alpha\beta\alpha}/\rho_M}$ denoting the velocity of the compressional $(\alpha = \beta)$ and shear $(\alpha \neq \beta)$ waves along direction $\mathbf{e}_\beta$ in the classical equivalent continuum ($S_{\alpha\beta\beta\alpha\beta\beta} = J_{\alpha\alpha\beta\beta} = 0$ in equation (12)), $\lambda_\beta^\alpha = \sqrt{S_{\alpha\beta\beta\alpha\beta\beta}/C_{\beta\alpha\beta\alpha}}$ the extensional $(\alpha = \beta)$ and shearing $(\alpha \neq \beta)$ characteristic lengths of the heterogeneous material, $J_{\alpha\beta\alpha\beta} = I_{\alpha\beta\alpha\beta} \rho_M \varepsilon^2$ the second-order inertia tensor ($I_{\alpha\alpha\beta\beta}$ being fourth-order tensor depending on the geometrical and mechanical properties of the cell) and $J_{\alpha\beta\beta\alpha\beta\beta} = I_{\alpha\beta\beta\alpha\beta\beta} \rho_M \varepsilon^4$ the fourth-order inertia tensor ($I_{\alpha\beta\beta\alpha\beta\beta}$ being sixth-order tensor depending on the geometrical and mechanical properties of the cell).

In order to obtain the dispersion functions, let us seek solutions to equation (13) of the form $U_\alpha(x_\beta, t) = A \exp\left[i(kx_\beta - \omega t)\right]$, where $i^2 = -1$, $k$ is the wave number and $\omega$ is the angular frequency. The wavelength and the phase velocity of the in-plane waves along direction $\mathbf{e}_\beta$ are $\lambda = 2\pi/k$ and $c_\beta^\alpha = \omega/k$, respectively. The dispersion function in terms of phase velocity corresponding to both the longitudinal $(\alpha = \beta)$ and the transverse $(\alpha \neq \beta)$ oscillatory motion in the second gradient equivalent continuum characterized by constants (9) and (10) assumes the following form

$$\frac{c_\beta^\alpha}{\hat{c}_\beta^\alpha} = \frac{\omega}{k\hat{c}_\beta^\alpha} = \sqrt{\frac{1+\left(k\lambda_\beta^\alpha\right)^2}{1+I_{\alpha\beta\alpha\beta}\left(k\varepsilon\right)^2+I_{\alpha\beta\beta\alpha\beta\beta}\left(k\varepsilon\right)^4}} = \sqrt{\frac{1+4\pi^2\left(\frac{\lambda_\beta^\alpha}{\lambda}\right)^2}{1+4\pi^2 I_{\alpha\beta\alpha\beta}\left(\frac{\varepsilon}{\lambda}\right)^2+16\pi^4 I_{\alpha\beta\beta\alpha\beta\beta}\left(\frac{\varepsilon}{\lambda}\right)^4}} \, . \quad (14)$$

Making the limit for $\lambda \to \infty$ (large wavelengths) of equation (14), the dispersion relation corresponding to a classical first order continuum is derived, i.e. $\frac{c_\beta^\alpha}{\hat{c}_\beta^\alpha} = \frac{\omega}{k\hat{c}_\beta^\alpha} \to 1$. If the third and fourth order inertial terms of are neglected in (8), the dispersion relation (14) takes the simplified form introduced in Bacigalupo and Gambarotta, 2012. For values of the dimensionless wavenumber $kd_\beta > \pi$ (with $\beta = 1,2$ and $d_1 = \varepsilon$, $d_2 = \delta\varepsilon$) a approximation of the dispersion curves may be obtained assuming that the solution of the equation of motion (12) has the form $U_\alpha\left(x_\beta,t\right) = A\exp\left[i\left(\left(k-2\pi n/d_\beta\right)x_\beta - \omega t\right)\right]$ (with $n \in \mathbb{Z}$, $n \geq 1$ and index $\beta$ not summed). In this case, for $\pi(2n-1) < kd_\beta < \pi(2n+1) \quad \forall n \in \mathbb{Z}$, one obtains the following dispersion relation

$$\frac{c_\beta^\alpha}{\hat{c}_\beta^\alpha} = \frac{\omega}{k\hat{c}_\beta^\alpha} = \left|1-\frac{2\pi n}{kd_\beta}\right|\sqrt{\frac{1+\left(k\varepsilon-2\pi n\varepsilon/d_\beta\right)^2\left(\lambda_\beta^\alpha/\varepsilon\right)^2}{1+I_{\alpha\beta\alpha\beta}\left(k\varepsilon-2\pi n\varepsilon/d_\beta\right)^2+I_{\alpha\beta\beta\alpha\beta\beta}\left(k\varepsilon-2\pi n\varepsilon/d_\beta\right)^4}} \, , \quad (15)$$

and for $n = 0$ this dispersion function takes the form of equation (14).

The dispersion function in terms of group velocity $c_{g\_\beta}^\alpha = \frac{d\omega}{dk}$ take the following form

$$\begin{aligned}\frac{c_{g\_\beta}^\alpha}{\hat{c}_\beta^\alpha} &= \frac{1+\left(k\lambda_\beta^\alpha\right)^2\left(2+I_{\alpha\beta\alpha\beta}\left(k\varepsilon\right)^2\right)-I_{\alpha\beta\beta\alpha\beta\beta}\left(k\varepsilon\right)^4}{\left(1+I_{\alpha\beta\alpha\beta}\left(k\varepsilon\right)^2+I_{\alpha\beta\beta\alpha\beta\beta}\left(k\varepsilon\right)^4\right)^2}\left(\frac{c_\beta^\alpha}{\hat{c}_\beta^\alpha}\right)^{-1} = \\ &= \frac{1+4\pi^2\left(\frac{\lambda_\beta^\alpha}{\lambda}\right)^2\left(2+4\pi^2 I_{\alpha\beta\alpha\beta}\left(\frac{\varepsilon}{\lambda}\right)^2\right)-16\pi^4 I_{\alpha\beta\beta\alpha\beta\beta}\left(\frac{\varepsilon}{\lambda}\right)^4}{\left(1+4\pi^2 I_{\alpha\beta\alpha\beta}\left(\frac{\varepsilon}{\lambda}\right)^2+16\pi^4 I_{\alpha\beta\beta\alpha\beta\beta}\left(\frac{\varepsilon}{\lambda}\right)^4\right)^2}\left(\frac{c_\beta^\alpha}{\hat{c}_\beta^\alpha}\right)^{-1} \, ,\end{aligned} \quad (16)$$

where $\frac{c_\beta^\alpha}{\hat{c}_\beta^\alpha}$ is the normalize phase velocity introduced in relation (14). If the third and fourth order inertial terms of are neglected in (8), the dispersion relation (14) takes the form obtained by *simplified dynamic homogenization procedure – SDHP* in Bacigalupo and Gambarotta, 2012. In this simplified approach the terms in $\varepsilon^4$ are neglected in dispersion relations associated to both normalize phase $\dfrac{c_\beta^\alpha}{\hat{c}_\beta^\alpha}$ and group velocities $\dfrac{c_{g\_\beta}^\alpha}{\hat{c}_\beta^\alpha}$, equations (14), (15) and (16), respectively.

Overall elastic moduli and inertial terms corresponding to periodic material shown in Figure 2 have been explicitly derived by means of second order dynamic homogenization technique. In next Section, these results are used together with dispersion relations (15), (16) and (17) in order to study acoustic waves motion in SOFC-like device.

## 4. Acoustic waves motion in planar SOFCs-like devices

Let us consider solid oxide fuel cell (SOFC) having periodic micro-structure as shown in Figure 2.a. and characterized by the periodic cell having characteristic size $\varepsilon = 100$ μm and $\delta = 4.4$ (see *Cell A* shown in Figure 4.a). The constituents are assumed to be isotropic, perfectly bonded and in plane strain condition. For temperature $T = 950$ °C the Young's moduli and Poisson ratios are assumed $E_1 = 155$ GPa, $\nu_1 = 0.3$, $E_2 = 50$ GPa, $\nu_2 = 0.25$, $E_3 = 130$ GPa, $\nu_3 = 0.3$, respectively. While the values $E_1 = 195$ GPa, $\nu_1 = 0.3$, $E_2 = 60$ GPa, $\nu_2 = 0.25$, $E_3 = 201$ GPa, $\nu_3 = 0.3$, correspond to temperature $T \simeq 21$ °C (see Kuebler *et al.*, 2010). The mass density for all the components is assumed $\rho_1 = \rho_2 = \rho_3 = 7$ g/cm$^3$.

The *cell problems* (3) have been numerically solved to obtain the micro-fluctuation functions $N_{hpq}^1(\xi)$ and $N_{hpqr}^2(\xi)$ by using the standard FEM solver in the Structural Mechanics Module, COMSOL 4.3 (COMSOL Multiphysics®, 2012). As mentioned in the Section 3, the periodicity boundary conditions on the micro-fluctuation functions have been prescribed for each problem. In both the cell problems, the jumps in the pseudo-body forces may take place at the interfaces between the inhomogeneities because of the mismatch between the elastic moduli. In this case the normal derivative across the interface takes the form of a Dirac–delta function which is modelled in the weak formulation and in the FE model through pseudo-forces applied at the interfaces with intensity equal to the above mentioned jump.

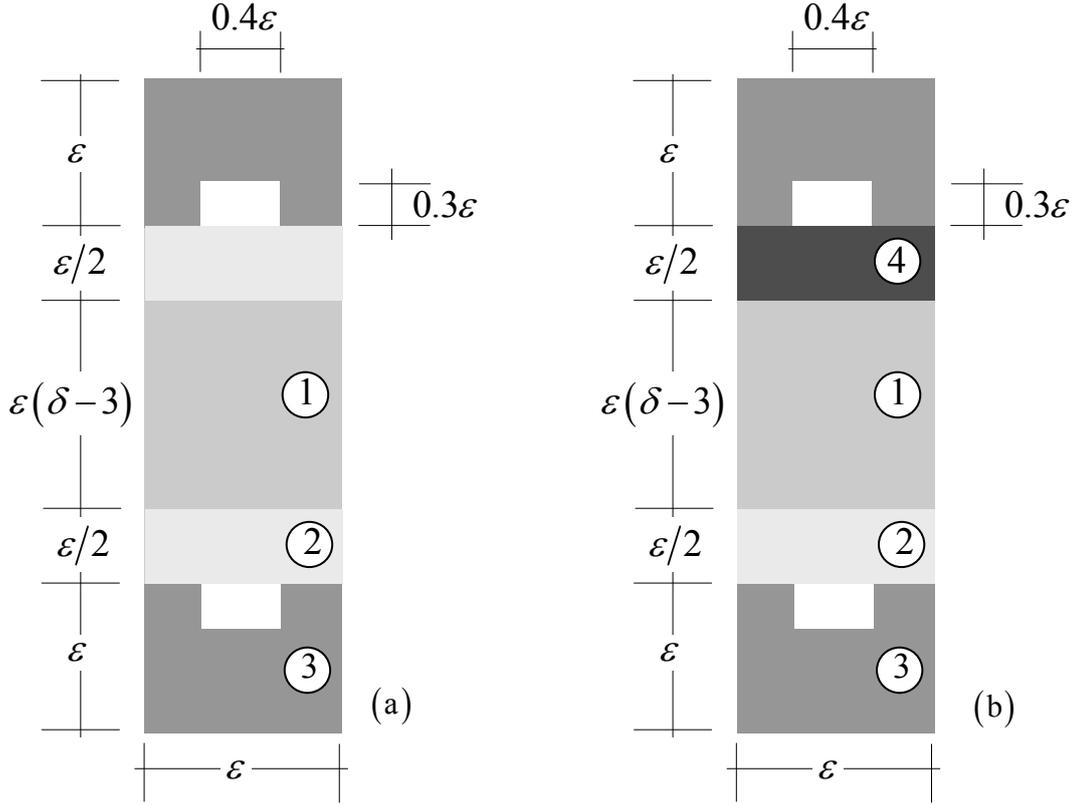

Figure 4. Periodic cell for the SOFC-like: (a) Cell A; (b) Cell B.

To estimate the reliability and the validity limits of the non-local dynamic homogenization method here proposed, the dispersion functions (15), (16) and (17) have been compared with those obtained by means of the rigorous Floquet-Bloch approach. The derivation of the acoustic waves dispersion relations by means of rigorous Floquet-Bloch procedure is reported in Appendix B.

The comparison between the two alternative approaches is carried out in terms of the dispersion relations and more precisely in terms of the ratio $\dfrac{c_\beta^\alpha}{\hat{c}_\beta^\alpha} = \dfrac{\omega}{k\hat{c}_\beta^\alpha}$. This comparison is based on the consideration (see Nemat-Nasser *et al*, 2011) that a homogenized material model is expected to describe the dynamic behaviour of the composite material provided that the overall conservation laws and compatibility relations (here satisfied by definition of the model), and the composite's dispersion relations are ensured.

The characteristic lengths $\lambda_\beta^\alpha = \sqrt{S_{\alpha\beta\beta\alpha\beta\beta}/C_{\beta\alpha\beta\alpha}}$ of the second gradient homogenized continuum and the waves velocities $\hat{c}_\beta^\alpha = \sqrt{C_{\beta\alpha\beta\alpha}/\rho_M}$ in the first-order homogeneous continuum are given in terms of temperature $T$ in Table 1 and Table 2, respectively. The overall mass density is $\rho_M = 7 \text{ g/cm}^3$ and the non-vanishing components $I_{\alpha\beta\alpha\beta}$ the micro-inertia tensors needed to represent the compressional and shear waves along the orthotropy axes are given in Table 3.

Table 1: Characteristic lengths $\lambda_\beta^\alpha$ (μm).

| $T\ (^\circ C)$ | $\lambda_{Sh-1} = \lambda_1^2$ | $\lambda_{Ext-1} = \lambda_1^1$ | $\lambda_{Sh-2} = \lambda_2^1$ | $\lambda_{Ext-2} = \lambda_2^2$ |
|---|---|---|---|---|
| 950 | 39.6 | 6.37 | 6.02 | 2.21 |
| 21 | 40.6 | 6.71 | 5.87 | 2.04 |

Table 2: Waves velocity $\hat{c}_\beta^\alpha$ (m/s).

| $T\ (^\circ C)$ | $\hat{c}_1^2 = \hat{c}_2^1$ | $\hat{c}_1^1$ | $\hat{c}_2^2$ |
|---|---|---|---|
| 950 | 2113.59 | 4414.89 | 3864.45 |
| 21 | 2447.08 | 5158.28 | 4437.57 |

Table 3: The non-vanishing components $I_{\alpha\beta\alpha\beta}$.

| $T\ (^\circ C)$ | $I_{2121} = I_{1212}$ | $I_{1111}$ | $I_{2222}$ |
|---|---|---|---|
| 950 | $4.265 \times 10^{-2}$ | $1.202 \times 10^{-2}$ | $4.848 \times 10^{-2}$ |
| 21 | $4.156 \times 10^{-2}$ | $1.267 \times 10^{-2}$ | $5.273 \times 10^{-2}$ |

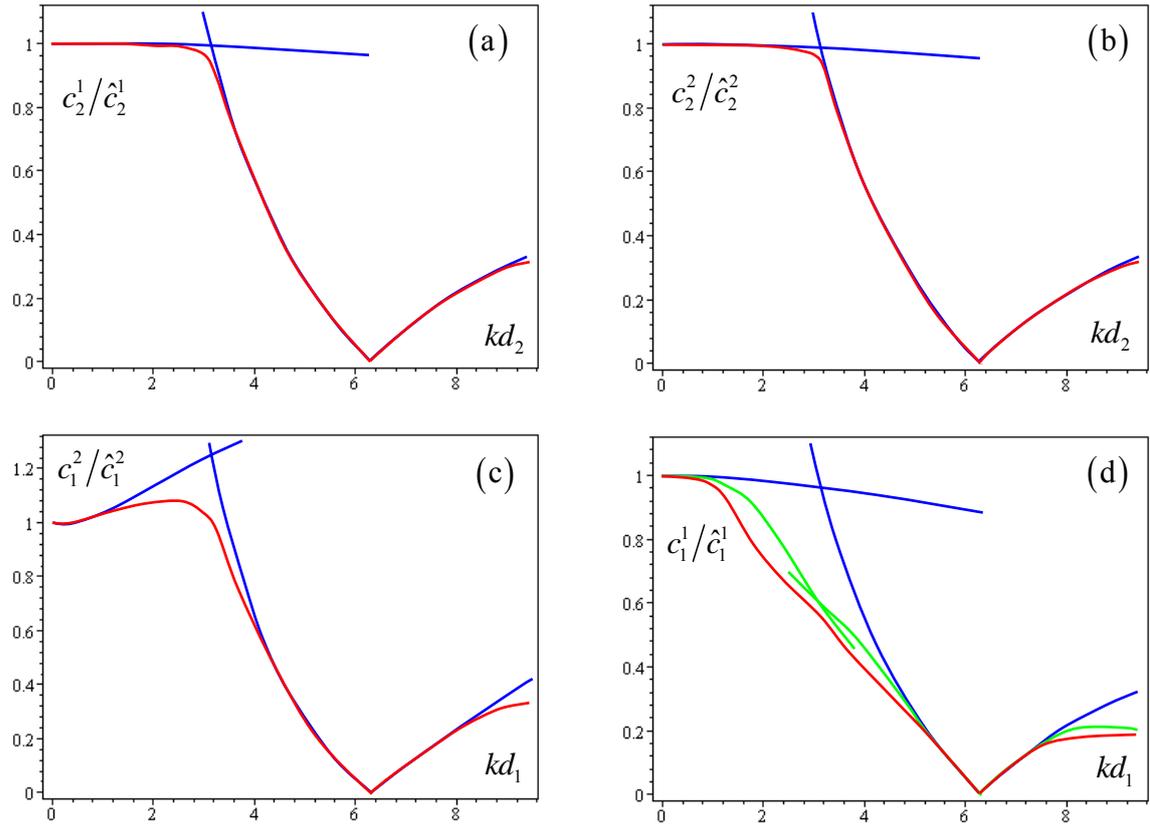

Figure 5: Shear and compressional waves for temperature $T = 950\ °C$: normalized phase velocities versus normalized wavenumber. Red line: heterogeneous material - Floquet-Bloch approach; Blue line: second-gradient continuum - *SDHP*; Green line: second-gradient continuum. (a) S. waves in $e_2$ direction; (b) C. waves in $e_2$ direction; (c) S. waves in $e_1$ direction; (d) C. waves in $e_1$ direction.

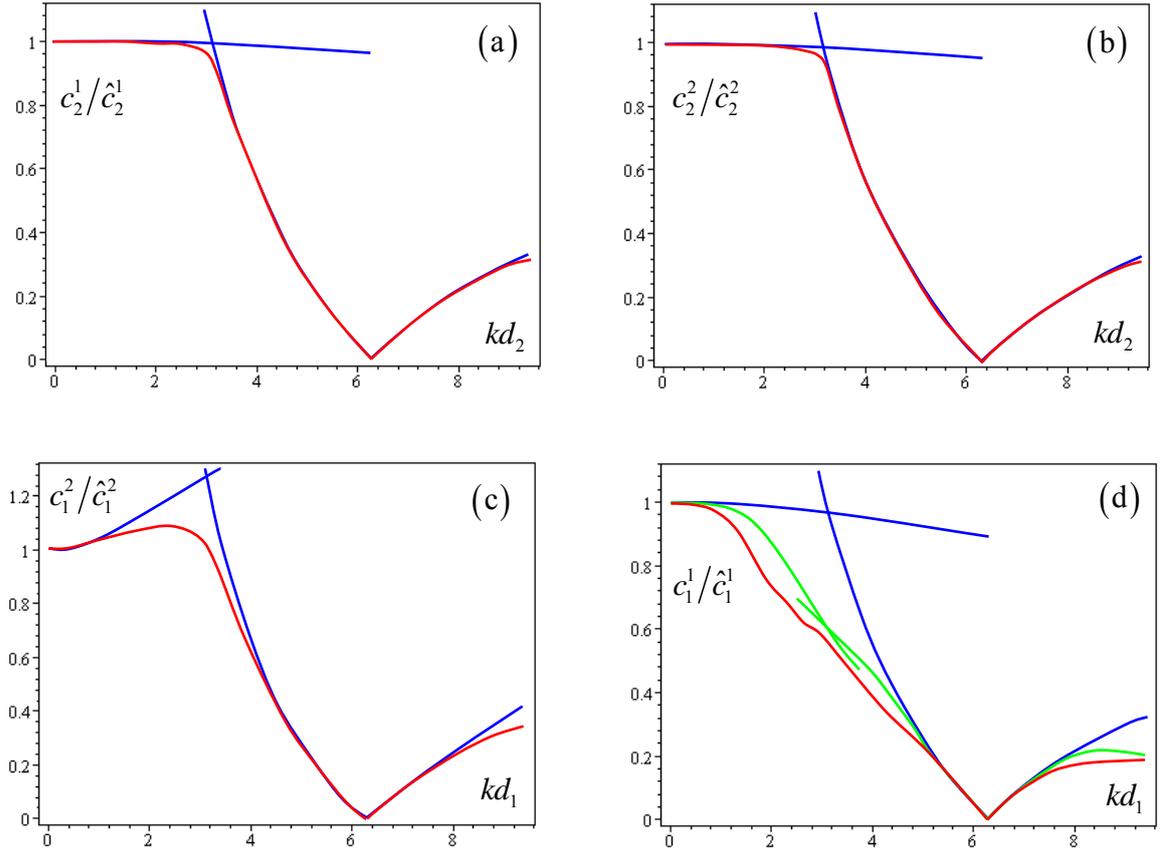

Figure 6: Shear and compressional waves for temperature $T = 21$ °C: normalized phase velocities versus normalized wavenumber. Red line: heterogeneous material - Floquet-Bloch approach ; Blue line: second-gradient continuum - *SDHP*; Green line: second-gradient continuum. (a) S. waves in $e_2$ direction; (b) C. waves in $e_2$ direction;   (c) S. waves in $e_1$ direction; (d) C. waves in $e_1$ direction.

The variation of normalized phase velocity $c_\beta^\alpha / \hat{c}_\beta^\alpha$ of both shear and compressional waves is shown in the diagrams of Figure 5 and 6 for $T = 950$ °C and $T = 21$ °C, respectively, as a function of the normalized wavenumber $kd_\beta$ (with $\beta = 1, 2$ and $d_1 = \varepsilon$, $d_2 = \delta \varepsilon$). The dispersion curves corresponding to shear and compressional waves along the directions of the layers and normal to the layers have been reported.

It is important to observe that the curves obtained by *SDHP* that disregards the term $\varepsilon^4$ (blue line) in equations (14) and (15) are in good agreement with those obtained by the Floquet-Bloch approach (red line). In agreement to what has been detected in Bacigalupo and Gambarotta, 2013, 2014) this model seems to allow sufficiently accurate results in cases where the cell size is limited. Conversely, in layered bi-materials, the contributions fourth -order inertia tensor $I_{\alpha\beta\beta\alpha\beta\beta}$ (sixth-order tensor) is relevant and then not negligible for studying compressional waves propagation along the layers (Bacigalupo and Gambarotta, 2014). For normalized wavenumber $kd_\beta > \pi$, a good approximation of the dispersion curves is obtained by equation (15) for $\pi(2n-1) < kd_\beta < \pi(2n+1) \quad \forall n \in \mathbb{Z}$. This can be observed in Figure 5 and 6 where the dispersion curves associated to $n = 0$ and $n = 1$ are plotted, respectively. It is important to note that, as expected, the dispersion functions (14) and (15) associated with compressional waves along the directions of the layers obtained by the dynamic homogenization procedure (green line) provide a closer approximation of the exact results obtained by the Floquet-Bloch theory (red line) respect to that derived by *simplified dynamic homogenization procedure* (*SDHP*).

The normalized group velocity $c^\alpha_{g\_\beta}/\hat{c}^\alpha_\beta$ (black line and violet line for *SDHP*) and phase velocity $c^\alpha_\beta/\hat{c}^\alpha_\beta$ (green line and blue line for *SDHP*) of both shear and compressional waves are shown in Figures 7 for temperature $T = 950$ °C and $T = 21$ °C, respectively, in terms of the normalized wave-length $\lambda/d_\beta \geq 2$ (with $\beta = 1,2$ and $d_1 = \varepsilon$, $d_2 = \delta\varepsilon$). From the diagrams of Figure 7 it emerges that the difference between the velocity of dispersive waves along the layers and the corresponding one in the classical continuum is much higher compared to the corresponding results referred to the propagation along the normal direction to the layers. This result is in agreement with the variation of the characteristic lengths detected in Table 1. Indeed, in this Table one can observe that the characteristic lengths along the layer are greater than those associated to the normal direction. Furthermore, in direction normal to the layer, both the phase and group velocities differ very little from the ones obtained by considering the classical continuum and reported in Table 2. In Figure 7 it is shown that the acoustic branch of the Bloch spectrum is weakly dependent by the temperature $T$. Indeed, the behaviour of both the phase and group velocity in the second gradient continuum obtained for $T = 21$ °C (linespoints) is very similar to those derived for $T = 950$ °C.

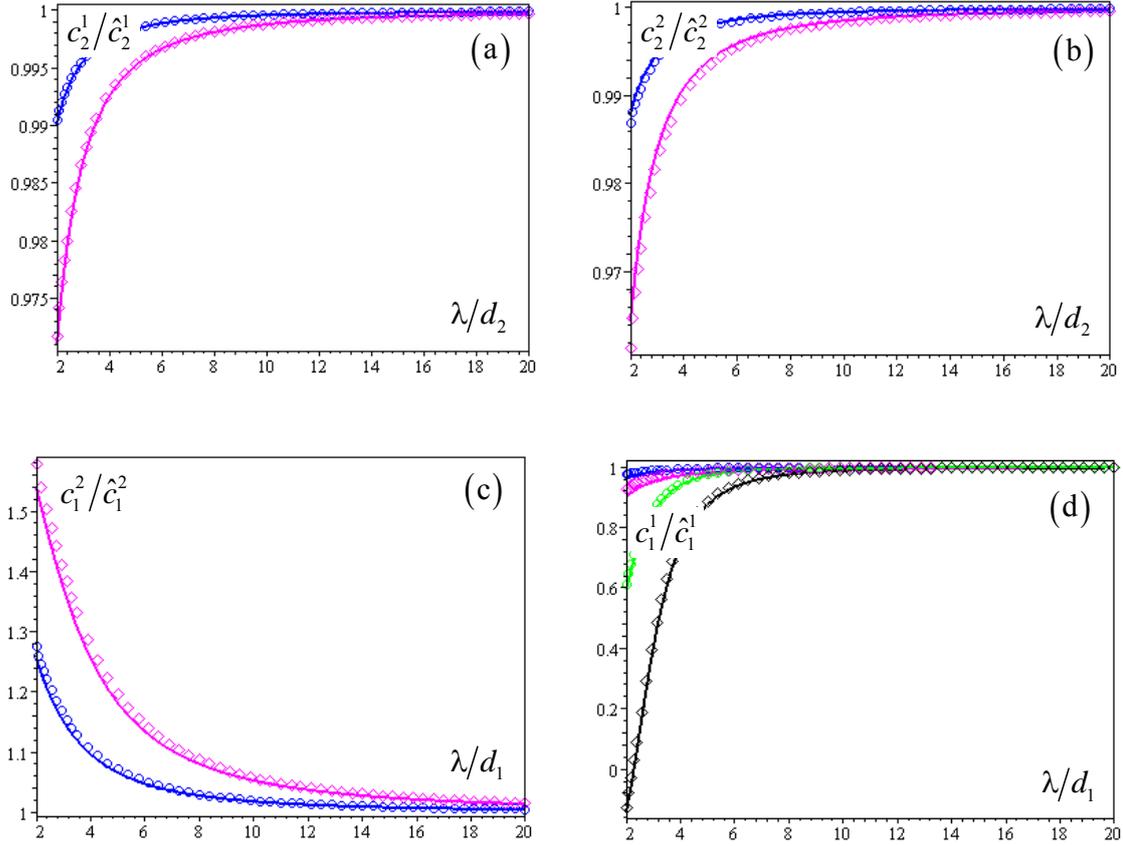

Figure 7: Shear and compressional waves in homogenized second gradient continuum: normalized phase and group velocities versus non-dimensional waveleght. Phase velocity: green line (blue line - *SDHP*) for $T = 950$ °C; green linespoints (blue linespoints - *SDHP*) for $T = 21$ °C. Group velocity: black line (violet line - *SDHP*) for $T = 950$ °C; black linespoints (violet linespoints - *SDHP*) for $T = 21$ °C. (a) S. waves in $e_2$ direction; (b) C. waves in $e_2$ direction; (c) S. waves in $e_1$ direction; (d) C. waves in $e_1$ direction.

The results reported in Figures 5 and 6 show that the dynamic homogenization procedure introduced in Section 3 provides a sufficiently accurate approximation of the lowest (acoustic) branch of the Bloch spectrum for a wide range of wavelengths.

Differently, the high frequencies range can be studied by means of FEM implementation of the Floquet-Bloch procedure described in Appendix B. In order to detect the presence of pass and stop frequency bands, full Bloch spectrum is determined and investigated. The normalized angular frequency $\omega d_\beta / 2\pi \hat{c}_\beta^\alpha$ in terms of the normalized wavenumber $k d_\beta$ (with $\beta = 1, 2$ and $d_1 = \varepsilon$, $d_2 = \delta\varepsilon$) are shown in the diagrams of Figure 8 for $T = 950$ °C (red line) and $T = 21$ °C (cyan dash-dot line). The dispersion curves associated to the lower four branches of propagation are plotted for both shear and compressional waves travelling respectively along the directions of the layers and normal to the layers. The presence of Band-gaps is observed for both shear and compressional waves in the direction normal to the layer (Figure 8.a e 8.b). Conversely, band-gaps are not detected in the case where the propagation is directed along the layer (Figure 8.c and 8.d). These results are in agreement with those normally obtained in studies regarding waves propagation in three-layered materials. This means that the presence of voids associated to fuel/air channels in the considered bi-dimensional SOFC-like structure do not influence significantly the qualitative behaviour of the Bloch spectrum. As it can be observed in Figure 8, the temperature affect slightly the high-frequency branches of the Bloch spectrum. The band-gaps amplitude increases with the temperature. This is due to the fact that the velocity of propagation of both shear and compressional waves in first order homogeneous continuum decreases with an increasing of the temperature.

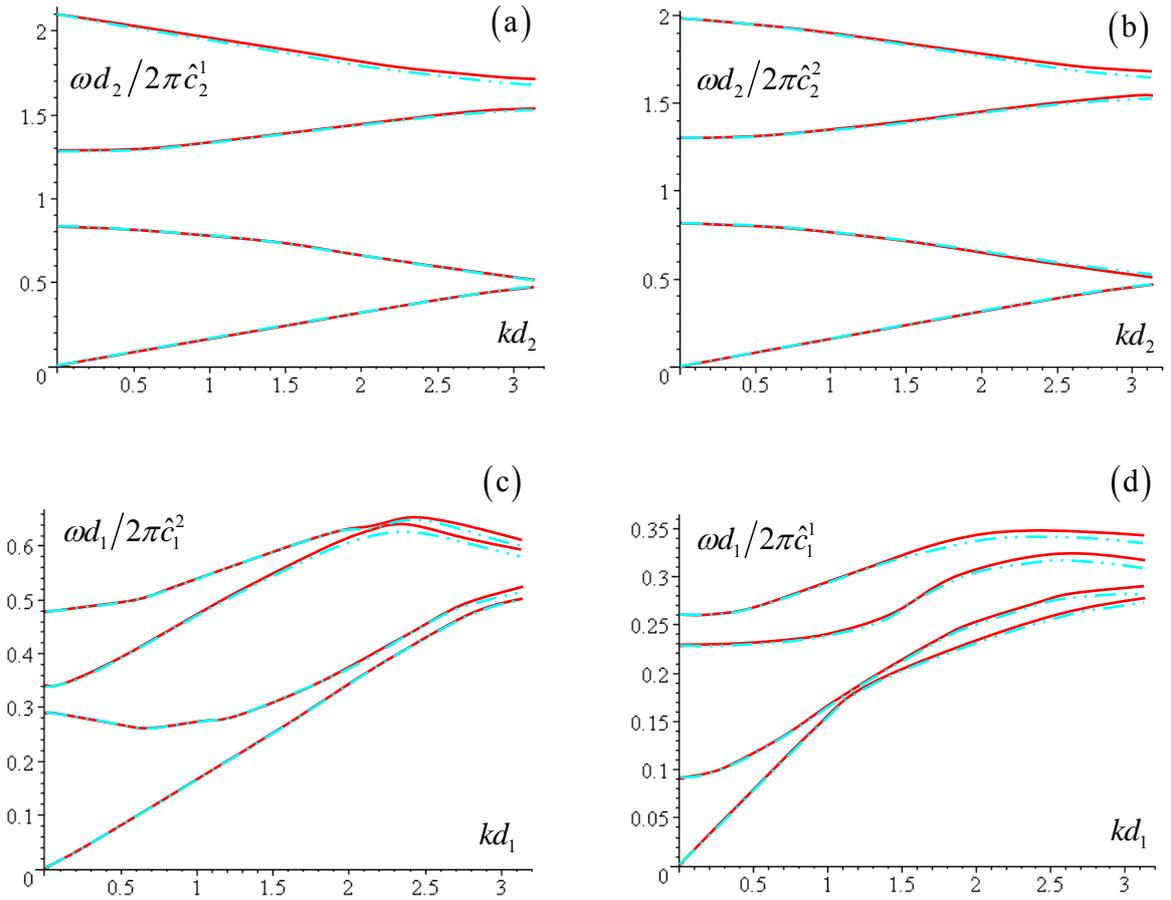

Figure 8: Shear and compressional waves for $T = 950$ °C (red line) and $T = 21$ °C (cyan dash-dot line): dispersion curves in the Brillouin zone. (a) S. waves in $e_2$ direction; (b) C. waves in $e_2$ direction; (c) S. waves in $e_1$ direction; (d) C. waves in $e_1$ direction.

As further example, let us consider solid oxide fuel cell (SOFC) characterized by the periodic cell having characteristic size $\varepsilon = 100$ μm and $\delta = 4.4$ shown in Figure 4.b. In order to test the influence of different elastic moduli of the cell elements on Bloch spectrum, the anode and the cathode elements are assumed to be realized with different materials. The constituents are assumed to be isotropic, perfectly bonded and in plane strain condition, the Young's moduli and Poisson ratios are given by $E_1 = 155$ GPa, $\nu_1 = 0.3$, $E_2 = 40$ GPa, $\nu_2 = 0.25$, $E_3 = 130$ GPa, $\nu_3 = 0.3$, $E_4 = 60$ GPa, $\nu_4 = 0.25$, respectively for temperature $T = 950$ °C (see Kuebler *et al.*, 2010). The mass density for the components is assumed $\rho_1 = \rho_2 = \rho_3 = \rho_4 = 7$ g/cm$^3$.

The velocity of the waves along direction **m** in the classical equivalent continuum is derived solving the following eigenvalue problem (Bedford and Drumheller, 1994)

$$\begin{bmatrix} Q_{11}^2 - \langle \rho \rangle c^2 & Q_{12}^2 \\ Q_{21}^2 & Q_{22}^2 - \langle \rho \rangle c^2 \end{bmatrix} \begin{Bmatrix} A_1 \\ A_2 \end{Bmatrix} = \begin{Bmatrix} 0 \\ 0 \end{Bmatrix} \qquad (17)$$

$Q_{ip}^2 = C_{ir_1pr_2} m_{r_1} m_{r_2}$ being the component of the acoustical first order tensor, $C_{ir_1pr_2}$ the components of the classical overall elastic fourth order tensor (equation (9.1)) and $c$ the eigenvalue having the meaning of wave phase velocity. For any wave direction vector **m**, the corresponding eigenvalues $c_\varsigma^2$ ($\varsigma = 1,2$) are related to the wave velocity $\hat{c}_\varsigma$ in the first order homogenized continuum; the associated eigenvectors (of components $A_p$) identify the directions of polarization.

In order to study the pass and stop frequency bands, the Bloch spectrum corresponding to the periodic material reported in Figure 4 is fully obtained and analysed. The normalized angular frequency $\omega d_\beta / 2\pi \hat{c}_\beta^*$ in terms of the normalized wavenumber $kd_\beta$ (with $\beta = 1,2$ and $d_1 = \varepsilon$, $d_2 = \delta \varepsilon$) in Figure 9 (with $T = 950$ °C) per both *Cell A* (red line for symmetric and blue line for anti-symmetric mode) and *Cell B* (green dash-dot line). With the purpose of normalize the dispersion curves, the reference velocity value $\hat{c}_\beta^* = \frac{1}{2}(\hat{c}_{1\_\beta} + \hat{c}_{2\_\beta})$ is introduced. Where $\hat{c}_{1\_\beta}$, $\hat{c}_{2\_\beta}$ are related to the eigenvalues $c_\varsigma^2$ ($\varsigma = 1,2$) solution of the eigenvalue problem (17) for $\mathbf{m} = \mathbf{e}_\beta$. Assuming propagation along the direction of $\mathbf{e}_2$, the eigenvalues are given by: $\hat{c}_{1\_2} = 2096.62$ m/s, $\hat{c}_{2\_2} = 3826.61$ m/s for *Cell B* and $\hat{c}_{1\_2} = 2113.59$ m/s, $\hat{c}_{2\_2} = 3864.45$ m/s (see Tab. 2) for *Cell A*. In this case, the reference velocity $\hat{c}^*_\beta$ is: $\hat{c}^*_2 = 2961.61$ m/s for *Cell B* and $\hat{c}^*_2 = 2989.02$ m/s for *Cell A*. If the waves propagate along of $\mathbf{e}_1$ the results are: $\hat{c}_{1\_1} = 2096.62$ m/s, $\hat{c}_{2\_1} = 4409.85$ m/s, $\hat{c}^*_1 = 3253.23$ m/s for *Cell B* and $\hat{c}_{1\_1} = 2113.59$ m/s, $\hat{c}_{2\_1} = 4414.89$ m/s (see Tab. 2), $\hat{c}^*_2 = 3264.24$ m/s for *Cell A*.

In Figure 9 some differences between the Bloch spectra corresponding to *Cell A* and *Cell B* can be noted in the high frequencies branches. At low frequencies, the variation of elastic moduli of cathode and anode constituents ($|E_4 - E_2| \approx 20$ GPa), do not induce significative differences in the corresponding dispersion curves.

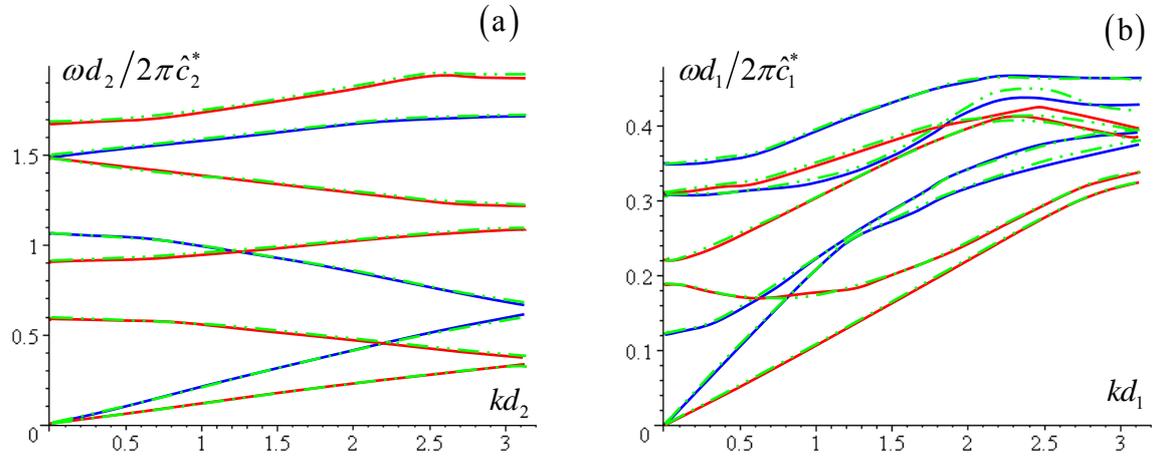

Figure 9: Dispersion curves in the Brillouin zone: normalized angular frequency $\omega d_\beta / 2\pi \hat{c}^*_\beta$ v.s. normalized wavenumber $kd_\beta$ for *Cell A* (red line S. waves and blue line C. waves) and for *Cell B* (green dash-dot line). (a) waves in $e_2$ direction; (b) waves in $e_1$ direction.

## 5. Conclusions

Exact expressions for the overall elastic moduli and inertial terms of the second gradient continuum equivalent to multi-layered SOFC-like materials have been derived by means of dynamic homogenization procedure. The proposed approach allows a synthetic description of the size effects on mechanical properties of SOFC devices, avoiding the general local solution of dynamic problems in this kind of heterogeneous material and the connected challenging computational problems. The equivalent non-local continuum here defined by the homogenization method is particularly appropriate for modelling the SOFC's response in presence of strong stress gradients.

The derived formulae for the overall elastic and inertial constants have been used together with analytical dispersion relations for studying acoustic waves propagation in non-local continuum media equivalent to periodic SOFC-like devices. Both shear and compressional acoustic waves have been considered, and the relative phase and group velocities in second gradient homogeneous media have been studied. The obtained results are in good agreement with those derived by the rigorous Floquet-Bloch approach.

From the waves propagation analysis it emerges that the difference between the velocity of dispersive waves along the layers and the corresponding one in the classical continuum is much higher compared to the corresponding results referred to the propagation along the normal direction to the layers. This means that the non-local effects quantified by the characteristic lengths are more relevant along the layer direction, whereas they are negligible in normal direction of the layer. The values of both the phase and group velocity is detected to be weakly dependent by the temperature of the media.

High frequencies propagation branches are investigated by means of Floquet-Bloch approach. Band-gaps are detected for both shear and compressional waves propagating along the direction normal to the layer. Conversely, band-gaps are not detected in the case where the propagation is directed along the layer. This result is analogous to what is commonly observed in stratified materials without voids. It means that the presence of voids associated to fuel/air channels in the considered bi-dimensional SOFC-like devices do not influence significantly the qualitative behaviour of the Bloch spectrum. The analysis show that the difference between the values of the angular frequency belonging to adjacent dispersive branches increases with the temperature. This can be explained with the fact that the velocity of propagation of both shear and compessional waves in first order homogeneous continuum decreases with an increasing of the temperature.

The influence of lacking of centro-symmetry in the elastic properties of the periodic cell on the Bloch spectrum has been analyzed. The variation of elastic moduli of cathode and anode constituents do not influence significantly the dispersion relation of the waves in terms of normalized angular frequency.

The exact expressions obtained for the overall elastic moduli and inertial terms of non-local homogeneous continuum equivalent to periodic SOFC-like material have been tested by means of acoustic wave propagation analysis. This explicit expressions may represent a useful tool in order to optimize elastic properties for the design and realization of planar SOFC energy devices.

**Acknowledgements**

The authors gratefully thank financial support of the Italian Ministry of Education, University and Research in the framework of the FIRB project 2010, "Structural mechanics models for renewable energy applications".

**References**

Addessi D., De Bellis M.L., Sacco E., Micromechanical analysis of heterogeneous materials subjected to overall Cosserat strains, *Mechanics Research Communications*, 54, 27-34, 2013.

Atkinson A., Sun B. Residual stress and thermal cycling of planar solid oxide fuel cells, *Materials Science and Technology*, **23** (10),1135–43, 2007.

Bakhvalov N.S., Panasenko G.P., Homogenization: Averaging Processes in Periodic Media. Nauka, Moscow (in Russian). English translation in: Mathematics and its Applications (Soviet Series) 36, Kluwer Academic Publishers, Dordrecht-Boston-London, 1984.

Bacca, M., Bigoni, D., Dal Corso, F. and Veber, D., Mindlin second-gradient elastic properties from dilute two-phase Cauchy-elastic composites. Part I: Closed form expression for the effective higher-order constitutive tensor, *Int. J. of Solids and Structures*, **50**, 4010-4019, 2013a.

Bacca, M., Bigoni, D., Dal Corso, F. and Veber, D., Mindlin second-gradient elastic properties from dilute two-phase Cauchy-elastic composites Part II: Higher-order constitutive properties and application cases, *International Journal of Solids and Structures*, **50**, 4020-4029, 2013b.

Bacca M., Dal Corso F. Veber D. and Bigoni D. Anisotropic effective higher-order response of heterogeneous Cauchy elastic materials, *Mechanics Research Communications*, **54**, 63-71, 2013c.

Bacigalupo A., Gambarotta L., Computational two-scale homogenization of periodic masonry: characteristic lengths and dispersive waves, *Computer Methods in Applied Mechanics and Engineering*, **213–216**, 16–28, 2012.

Bacigalupo A., Gambarotta L., Multi-scale strain-localization analysis of a layered strip with debonding interfaces, *International Journal of Solids and Structures*, **50**, 2061-2077, 2013a.

Bacigalupo A., Gambarotta L., Second-gradient homogenized model for wave propagation in heterogeneous periodic media, *International Journal of Solids and Structures*, **51**, 1052–1065, 2013b.

Bacigalupo A., Second-order homogenization of periodic materials based on asymptotic approximation of the strain energy: formulation and validity limits, *Meccanica*, DOI 10.1007/s11012-014-9906-0, 2014 (to appear).

Bacigalupo A., Gambarotta L., Computational dynamic homogenization for the analysis of dispersive waves in layered rock masses with periodic fractures, *Computers&Geotechnics*, **56**, 61-68, 2014.

Bedford A., Drumheller D.S., *Elastic Wave Propagation*, John Wiley & Sons, 1994.

Bove R., Ubertini S. (Eds.), Modeling Solid Oxide Fuel Cells: Methods, Procedures and Techniques, Springer, Netherlands, 2008.

Brandon N.P., Brett D.J., Engineering porous materials for fuel cell applications, *Phil. Trans R. Soc. A*, **364**, 147–159, 2006.

Colpan C.O., Dincer I., Hamdullahpur F., A review on macrolevel modeling of planar solid oxide fuel cells, *Int. J. Energy Res*. **32**, 336–355, 2008.

De Bellis M.L., Addessi D., A Cosserat based multi-scale model for masonry structures, *International Journal for Multiscale Computational Engineering*, **9**, 543–563, 2011.

Deseri L, Owen D.R., Toward a field theory for elastic bodies undergoing disarrangements, *J. Elast*., **70** (1), 197–236, 2003.

Deseri L., Owen D.R., Submacroscopically stable equilibria of elastic bodies undergoing disarrangements and dissipation, *Math. Mech. Solids*, **15** (6), 611–638, 2010.

Forest S., Sab K., Cosserat overall modeling of heterogeneous materials, *Mech. Res. Comm.*, **25**, 449-454, 1998.

Forest S., Trinh D.K., Generalised continua and non-homogeneous boundary conditions in homogenisation methods, *ZAMM Z. Angew. Math. Mech.*, **91**, No. 2, 90–109, 2011.

Hajimolana S.A., Hussain M.A., Wan Daud W.M.A., Soroush M., Shamiri A., Mathematical modeling of solid oxide fuel cells: a review, Renew. Sustain. Energy Rev. **15** (4), 1893–1917, 2011.

Kakaç S., Pramuanjaroenkij A., Zhou X.Y., A review of numerical modeling of solid oxide fuel cells, *Int. J. Hydrogen Energy*, **32**, 761–786, 2007.

Kanoute P., Boso D.P., Chaboche J.L., Schrefler B.A., Multiscale methods for composites: a review. Archives of Computational Methods in Engineering, **16**, 31–75, 2009.

Kim J.H., Liu W.K., Lee C. Multi-scale solid oxide fuel cell materials modeling. *Computational Mechanics*; **44** (5), 683–703, 2009.

Kuebler J., Vogt U.F., Haberstock D., Sfeir J., Mai A., Hocker T. Roos M., Simulation and validation of thermo-mechanical stresses in planar SOFCs, *Fuel Cells*, **10** (6), 1066-1073, 2010.

Mindlin R.D., Micro-structure in linear elasticity, *Arch. Ration. Mech. Anal.*, 16, 51–78, 1964.

Nemat-Nasser S., Willis J.R., Srivastava A., Amirkhizi A.V., Homogenization of periodic elastic composites and locally resonant sonic materials, Physical Review B, **83**, 104103, 2011.

Pitakthapanaphong S., Busso E.P., Finite element analysis of the fracture behaviour of multi layered systems used in solid oxide fuel cell applications. *Modelling and Simulation in Materials Science and Engineering*, **13** (4), 531-540, 2005.

Salvadori A., Bosco E., Grazioli D., A computational homogenization approach for Li-ion battery cells: Part1 − formulation, *Journal of the Mechanics and Physics of Solids*, **65**, 114–137, 2014.

Schrefler B.A., Boso D.P., Pesavento F., Gawin D., Marek L., Mathematical and numerical multi-scale modeling of multiphysics problems, Computer Assisted Mechanics and Engineering Sciences, **18**, 91–113, 2011.

Smyshlyaev V.P., Cherednichenko K.D., On rigorous derivation of strain gradient effects in the overall behaviour of periodic heterogeneous media, J. Mechanics and Physics of Solids, **48**, 1325–1357, 2000.

Sun C.T., Huang G.L., Modeling Heterogeneous Media with Microstructures of Different Scales, *J. of Applied Mechanics*, ASME, **74**, 203-209, 2007.

Wang Z.-P., Sun C.T., Modelling micro-inertia in heterogeneous materials under dynamic loading, *Wave Motion*, **36**, 473-485, 2002.

Zhang H.W., Zhang S., Bi J.Y., Schrefler B.A., Thermomechanical analysis of periodic multiphase materials by a multiscale asymptotic homogenization approach. Int J Numer Methods Eng, 69 (1), 87–113, 2007.

Zhu W.Z., Deevi S.C., A review on the status of anode materials for solid oxide fuel cells, *Mater. Sci. Eng. A*, **362**, 228–239, 2003.

**Appendix A**

The localization tensors $B_{ijpq}^{H}$, $B_{ijpqr}^{\kappa}$ and $A_{pqrjhk}^{H-\kappa}$ in the overall elastic moduli expressed in equation (9) take the form

$$B_{ijpq}^{H}\left(\xi = \frac{\mathbf{x}}{\varepsilon}\right) = \frac{1}{4}\left(\delta_{ip}\delta_{jq} + N_{ipq,j}^{1} + \delta_{jp}\delta_{iq} + N_{jpq,i}^{1} + \delta_{iq}\delta_{jp} + N_{iqp,j}^{1} + \delta_{jq}\delta_{ip} + N_{jqp,i}^{1}\right),$$

$$B_{ijpqr}^{\kappa}\left(\xi = \frac{\mathbf{x}}{\varepsilon}\right) = \frac{1}{4}\left(N_{ipq}^{1}\delta_{jr} + N_{ipr}^{1}\delta_{qj} + 2N_{ipqr,j}^{2} + N_{jpq}^{1}\delta_{ir} + N_{jpr}^{1}\delta_{qi} + 2N_{jpqr,i}^{2}\right),$$

$$\begin{aligned}A_{pqrjhk}^{H-\kappa}\left(\xi = \frac{\mathbf{x}}{\varepsilon}\right) &= \left(C_{jkih}^{m}N_{ipqr}^{2} + C_{jkir}^{m}N_{ipqh}^{2} + C_{jkiq}^{m}N_{iprh}^{2} + C_{jhik}^{m}N_{ipqr}^{2} + C_{jhir}^{m}N_{ipqk}^{2} + C_{jhiq}^{m}N_{iprk}^{2} + \right.\\
&+ C_{pqir}^{m}N_{ijkh}^{2} + C_{pqih}^{m}N_{ijkr}^{2} + C_{pqik}^{m}N_{ijhr}^{2} + C_{priq}^{m}N_{ijkh}^{2} + C_{prih}^{m}N_{ijkq}^{2} + C_{prik}^{m}N_{ijhq}^{2}\left.\right) + \\
&+ \frac{1}{2}\left(C_{stir}^{m}N_{ijkh}^{2}N_{spq,t}^{1} + C_{srit}^{m}N_{sjkh}^{2}N_{ipq,t}^{1} + C_{shit}^{m}N_{sjkr}^{2}N_{ipq,t}^{1} + C_{stik}^{m}N_{ijhr}^{2}N_{spq,t}^{1} + C_{skit}^{m}N_{sjhr}^{2}N_{ipq,t}^{1} + \right.\\
&+ C_{stiq}^{m}N_{ijkh}^{2}N_{spr,t}^{1} + C_{sqit}^{m}N_{sjkh}^{2}N_{ipr,t}^{1} + C_{stih}^{m}N_{ijkq}^{2}N_{spr,t}^{1} + C_{shit}^{m}N_{sjkq}^{2}N_{ipr,t}^{1} + C_{stik}^{m}N_{ijhq}^{2}N_{spr,t}^{1} + \\
&+ C_{skit}^{m}N_{sjhq}^{2}N_{ipr,t}^{1} + C_{stih}^{m}N_{ipqr}^{2}N_{sjk,t}^{1} + C_{shit}^{m}N_{spqr}^{2}N_{ijk,t}^{1} + C_{stir}^{m}N_{ipqh}^{2}N_{sjk,t}^{1} + C_{srit}^{m}N_{spqh}^{2}N_{ijk,t}^{1} + \\
&+ C_{stiq}^{m}N_{iprh}^{2}N_{sjk,t}^{1} + C_{sqit}^{m}N_{sprh}^{2}N_{ijk,t}^{1} + C_{stik}^{m}N_{ipqr}^{2}N_{sjh,t}^{1} + C_{skit}^{m}N_{spqr}^{2}N_{ijh,t}^{1} + C_{stir}^{m}N_{ipqk}^{2}N_{sjh,t}^{1} + \\
&+ C_{srit}^{m}N_{spqk}^{2}N_{ijh,t}^{1} + C_{stiq}^{m}N_{iprk}^{2}N_{sjh,t}^{1} + C_{sqit}^{m}N_{sprk}^{2}N_{ijh,t}^{1} + C_{stih}^{m}N_{ijkr}^{2}N_{spq,t}^{1}\left.\right).\end{aligned} \quad (18)$$

**Appendix B**

The dispersion relation are obtained by solving the spectral problem derived by the Fourier transform in the time variable $t$ of the microscopic motion equation (1)

$$div\left(\mathbb{C}^m\left(\frac{\mathbf{x}}{\varepsilon}\right)\nabla\hat{\mathbf{u}}(\mathbf{x})\right) + \omega^2\rho\left(\frac{\mathbf{x}}{\varepsilon}\right)\hat{\mathbf{u}}(\mathbf{x}) = \mathbf{0} \qquad (19)$$

with unknown the angular frequency ω and the displacement field $\hat{\mathbf{u}}(\mathbf{x})$ in the frequency space, respectively. To solve this spectral problem, the Floquet-Bloch boundary conditions have to be prescribed on the periodic cell $\mathscr{A} = [0,\varepsilon] \times [0,\delta\varepsilon]$ (Figure 4) in terms of the wave vector $\mathbf{k} = k\mathbf{m}$ ($k$ wave number, $\mathbf{m}$ unit vector of propagation) and of the periodicity vector $\mathbf{v}_p$, i.e.

$$\hat{\mathbf{u}}(\mathbf{x} + \mathbf{v}_p) = e^{i(\mathbf{k}\cdot\mathbf{v}_p)}\hat{\mathbf{u}}(\mathbf{x}) \; ,$$
$$\hat{\boldsymbol{\sigma}}(\mathbf{x} + \mathbf{v}_p)\mathbf{n}(\mathbf{x} + \mathbf{v}_p) = -e^{i(\mathbf{k}\cdot\mathbf{v}_p)}\hat{\boldsymbol{\sigma}}(\mathbf{x})\mathbf{n}(\mathbf{x}) \; , \qquad (20)$$

where $\mathbf{n}(\mathbf{x})$ is the outward normal unit vector at $\mathbf{x} \in \partial\mathscr{A}$ on the periodic cell boundary $\partial\mathscr{A}$ and $\hat{\boldsymbol{\sigma}}(\mathbf{x})$ is the stress tensor in the frequency space ($\hat{\boldsymbol{\sigma}} = \mathbb{C}^m\hat{\boldsymbol{\varepsilon}}$). The solution of the Floquet-Bloch problem is obtained by solving a FE model of the cell by the Structural Mechanics Module, COMSOL 4.3 (COMSOL Multiphysics®, 2012).